\documentclass[journal]{IEEEtran}

\usepackage{cite}
\usepackage{paralist}
\usepackage[cmex10]{amsmath}
\usepackage{amssymb, amsfonts}
\usepackage{graphicx}
\usepackage{epstopdf}
\usepackage{balance}
\usepackage{stfloats}
\usepackage{array}
\usepackage{bm}
\usepackage{color}
\usepackage{amsmath}
\usepackage{subfigure}
\usepackage{blkarray}
\usepackage{makecell}
\usepackage{booktabs}
\usepackage{soul}

\usepackage{mathrsfs}
\DeclareUnicodeCharacter{25C4}{\triangleleft}
\usepackage[font=footnotesize]{caption}
\IEEEoverridecommandlockouts  
\usepackage{url}             
\usepackage[
  paper=letterpaper,
  top=0.75in,
  bottom=1in,
  inner=0.625in,
  outer=0.65in,
  columnsep=0.22in
]{geometry}

\usepackage{algorithm,algorithmic}

\begin{document}

\title{Short Wins Long: Short Codes with Language Model Semantic Correction Outperform Long Codes\\
\thanks{Code available: https://github.com/Jeh100/SEC-for-Short-Block-Codes.git. The work of Chentao Yue was supported by ARC under Grant DE250101332.}
}

\author{
\IEEEauthorblockN{
Jiafu Hao,
Chentao Yue,
Hao Chang,
Branka Vucetic,
and Yonghui Li
}
\\
\IEEEauthorblockA{
School of Electrical and Computer Engineering, The University of Sydney, Australia\\
Email: \{jiafu.hao, chentao.yue, hao.chang, branka.vucetic, yonghui.li\}@sydney.edu.au
}
} 

\IEEEaftertitletext{\vspace{-2\baselineskip}}
\setlength{\abovedisplayskip}{2pt}
\setlength{\belowdisplayskip}{2pt}

\maketitle
\thispagestyle{empty}

\begin{abstract}
This paper presents a novel semantic-enhanced decoding scheme for transmitting natural language sentences with multiple short block codes over noisy wireless channels. After ASCII source coding, the natural language sentence message is divided into segments, where each is encoded with short block channel codes independently before transmission. At the receiver, each short block of codewords is decoded in parallel, followed by a semantic error correction (SEC) model to reconstruct corrupted segments semantically. We design and train the SEC model based on Bidirectional and Auto-Regressive Transformers (BART). Simulations demonstrate that the proposed scheme can significantly outperform encoding the sentence with one conventional long LDPC code, in terms of block error rate (BLER), semantic metrics, and decoding latency. Finally, we proposed a semantic hybrid automatic repeat request (HARQ) scheme to further enhance the error performance, which selectively requests retransmission depends on semantic uncertainty.
\end{abstract}


\vspace{-1em}
\section{Introduction}
\vspace{-0.3em}
\IEEEPARstart{U}{ltra}-reliable and low-latency communications (URLLC) is one of the key 5G service paradigms. 
The design of the physical layer, especially the channel coding scheme, for URLLC involves a fundamental trade-off between latency and reliability \cite{yue2023efficient}. While shorter code blocklengths can reduce transmission latency, they also significantly compromise coding gain and error-correction capability \cite{PPV}. Long low-density parity-check (LDPC) codes\cite{1057683} can approach the Shannon channel capacity and be efficiently decoded by belief propagation (BP) decoding \cite{Mahyar2019ShortCode}. However, their long blocklength introduce high propagation delay and processing latency at the receiver. Recently, short BCH code is shown to achieve the finite blocklength bound with the ordered-statistics decoding (OSD) \cite{yue2023efficient}, while their performance is inferior to long LDPC codes due to the channel distortion in the short blocklength regime \cite{PPV}.

Future 6G systems will prioritize intelligent and meaning-aware communication, emphasizing semantic fidelity over traditional bit-level accuracy \cite{9955525}. This paradigm shift positions semantic communication as a key technology for the next-generation networks. Current semantic communication research follows two primary directions. The first is joint source-channel coding (JSCC) that optimizes encoders and decoders jointly through end-to-end training \cite{jscc,deepsc,d2jscc}. 
The second focuses on semantic source coding, aiming to compress and transmit essential semantic content efficiently. For example, \cite{10494374} introduced importance-weighted semantic triples to identify key semantic information, while \cite{Lee} employed VQ-VAE for semantic compression to reduce transmission volume.

Despite these advances, semantic communication faces critical challenges for practical deployment in latency-sensitive 5G/6G applications. Most semantic approaches require a complete redesign of the transmitter and receiver with neural networks,  violating the established source–channel separation principle \cite{6773024} and incurring high implementation costs. Moreover, neural network-based JSCC often treats channel effects as abstract noise and learn to perform error correction directly during training, which overlooks decades of advances in channel coding theory and practice. Furthermore, semantic content (e.g., text, images, or video) typically requires  long codewords for transmission. Although these codewords approach Shannon capacity limits, they introduce delays that can conflict with strict latency requirements.

To tackle the aforementioned issues, we propose a novel receiver framework utilizing language models (LMs) for semantic error correction. Our framework uses multiple short block codewords to transmit one natural language sentence, and leverages contextual reasoning to reconstruct corrupted messages at the receiver. Multiple short blocks allow successfully decoded blocks to provide essential context for the LM to recover corrupted blocks. In contrast, long codeword transmission often results in a fully corrupted output upon decoding failure and loses most contextual information.

To enable contextual reasoning, we design and fine-tune a semantic error correction (SEC) model based on Bidirectional and Auto-Regressive Transformers (BART), particularly for the text recovery task. With SEC, our results show that transmitting 64-character sentences (i.e., 512 ASCII bits) using eight segments encoded with $(128,64)$ extended BCH codes significantly outperforms using a single $(1024,512)$ LDPC code with BP decoding. Compared to single LDPC, our approach achieves up to 0.5 dB gain in BLER at low SNRs, and acheives 1 dB gain in semantic fidelity as measured by BLEU \cite{bleu} and ROUGE-L \cite{rouge} score. Simultaneously, the decoding latency is reduced by approximately 50\%. Furthermore, we proposed the Semantic confidence (SemConf) HARQ to further enhance the  BLER and semantic fidelity performance, which selectively requests the retransmission of segments with the highest semantic uncertainty. Simulations show that even a single-round retransmission of one segment can improve both BLER and semantic fidelity by 0.3 dB.

\vspace{-0.6em}
\section{Background}
\vspace{-0.5em}
\label{section2}
\subsection{Linear block code} 
\label{section2a}
\vspace{-0.2em}
Linear block codes form the foundation of channel error correction in modern communication systems. A binary linear block code \(\mathcal{C}(n, k)\) transforms \(k\) information bits into an \(n\)-bit codeword, where \(n > k\), with the added redundancy providing error correction capability. These codes are defined by their  generator matrices \(\mathbf{G} \in \{0,1\}^{k \times n}\), and the encoding process maps a message vector \(\mathbf{b} \in \{0,1\}^k\) to a codeword through \(\mathbf{c} = \mathbf{bG} \in \{0,1\}^{n}\).

We assume that the codeword \(\mathbf{c}\) is modulated using binary phase shift keying (BPSK), resulting in \(\mathbf{x} = 1 - 2\mathbf{c} \in \{-1,1\}^n\). This signal is transmitted through an additive white Gaussian noise (AWGN) channel, and the received signal is given by
$\mathbf{y} = \mathbf{x} + \mathbf{z}$,
where \(\mathbf{z}=[z_1,\ldots,z_n] \) is the i.i.d. Gaussian noise vector, where each element follows \(z_i \sim \mathcal{N}(0, \sigma^2)\). The signal-to-noise ratio (SNR) is given by $\frac{1}{\sigma^2}$.

At the receiver, a decoder estimates the transmitted codeword as \(\hat{\mathbf{c}}\). A decoding error occurs if \(\hat{\mathbf{c}} \neq \mathbf{c}\). The decoder selection typically depends on the code structure and blocklength. 
For example, belief propagation (BP) decoding is a near-optimal decoding method for long LDPC codes \cite{Pearl1988-PEAPRI}, which performs iterative message passing on the code Tanner graph. On the other hand, ordered statistics decoding (OSD) is an effective near-optimal decoder for short block codes \cite{554278}, which approximates maximum likelihood decoding (MLD) with complexity $O(k^m)$, where $m = \lceil d_{\min} / 4 \rceil$ is the OSD order and $d_{\min}$ is the code minimum Hamming distance.

The best achievable BLER performance of block codes is fundamentally constrained by their blocklength. According to the normal approximation (NA) bound of the finite blocklength theory \cite{PPV}, for code $\mathcal{C}(n, k)$, its best BLER in binary AWGN channels is approximately given by \cite{erseghe2016coding}:

\begin{equation} \label{equ::PPV}
   \epsilon^*(k, n) \approx  Q\left( \sqrt{\frac{n}{V}}\cdot\left(\frac{C-R}{\log_2e}+\frac{\log n}{2n}\right)\right),
\end{equation}
where $C$ is the channel capacity, $V$ is the channel dispersion \cite[Fig. 6]{erseghe2016coding}, and $Q^{-1}(\cdot)$ is the inverse Gaussian Q-function. As shown by \eqref{equ::PPV}, at the same code rate $R$, the BLER performance of code $\mathcal{C}(n, k)$ significantly degrades as $n$ decreases. There is a fundamental trade-off: short block codes suffer higher error rates than longer codes, but long codes can introduce substantial delays, despite their better BLER performance. 

In this work, we will show that the error rate limitation of shot codes can be eliminated by semantic error correction.

\vspace{-0.8em}
\subsection{Pretrained Language Models}
\vspace{-0.3em}
\label{section2b} 
The Transformer is a neural network architecture initially introduced for natural language processing (NLP) tasks~\cite{NIPS2017_3f5ee243}. It models sequential data entirely through a self-attention mechanism and effectively captures contextual dependencies within the input. Building on this foundation, BART~\cite{bart} is a sequence-to-sequence model that integrates a bidirectional ``encoder'' and an autoregressive ''decoder''. Note that the encoder and decoder in BART refer to neural network components, which should be distinguished from the channel code encoder and decoder introduced earlier in Section \ref{section2a}.

The encoder in BART attends to the entire input sequence to produce rich contextual representations, while the decoder generates output tokens autoregressively, conditioned on both the encoder outputs and the previously generated tokens. BART is trained as a denoising autoencoder by corrupting the input text with various noise strategies (e.g., token masking, deletion, and sentence permutation), and minimizing a reconstruction loss. These strategies simulate text degradation, encouraging the model to learn robust, context-aware representations. Through this process, BART develops strong capabilities in semantic correction and sequence generation.

\begin{figure}[t]
\centering
\includegraphics[width=0.5\textwidth]{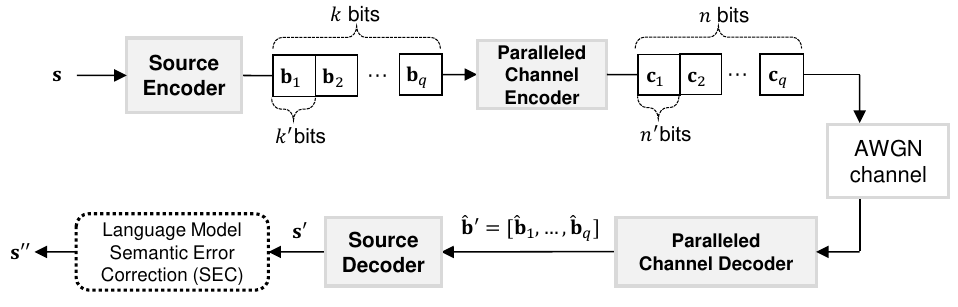}
\caption{Proposed MSC approach using multiple short block codes segments and SEC. The modulation and demodulation are omitted for clarity.}
\vspace{-1.1em}
\label{fig1}
\end{figure}

\vspace{-0.8em}
\section{Proposed Semantic Error Correction Scheme}
\label{section3}
\vspace{-0.4em}
\subsection{System Overview}
\label{section3a}
\vspace{-0.2em}

We consider the transmission of a natural language sentence $\mathbf{s}$ with character length $\ell$ over a noisy channel. For transmission, each character is first converted to its 8-bit ASCII representation, resulting in a binary bit steam $\mathbf{b}$ with $k=8\ell$ bits. We compare two distinct channel coding approaches:
\begin{itemize}
    \item Traditional Long Code (TLC): The entire sentence is encoded as a single unit. After ASCII conversion, the resulting bitstream is encoded by a $\mathcal{C}(n,k)$ LDPC code, which results in a single LDPC codeword \(\mathbf{c}\) of length $n$.
    \item Proposed Multiple Short Code (MSC): The sentence is divided into $q$ equal segments. After ASCII conversion, we obtain $q$ separate bitstreams $\mathbf{b}=\{\mathbf{b}_1,\ldots,\mathbf{b}_q\}$ each with length \(k' = \frac{k}{q}\). Then, each segment $\mathbf{b}_i$ is encoded independently using a $\mathcal{C}(n',k')$ BCH code, producing $q$ separate codewords $\{\mathbf{c}'_1,\mathbf{c}'_2,\ldots,\mathbf{c}'_q\}$, each with length $n_2$. At the receiver, after independent decoding of each segment, we apply semantic error correction (SEC) to leverage context from correctly decoded segments to recover corrupted ones. The process is illustrated in Fig.~\ref{fig1}.
\end{itemize}

Both schemes are designed to operate at the same code rate $R = \frac{k}{n} = \frac{k'}{n'}$. This implies $n' = {n}/{q}$, meaning the total number of coded bits remains the same for both methods.

For transmission, the $q$ codewords are concatenated to form $\mathbf{c}' = [\mathbf{c}'_1,\mathbf{c}'_2,\ldots, \mathbf{c}'_q]$ in our proposed MSC approach. We assume that the encoded codewords of both approaches, $\mathbf{c}$ and $\mathbf{c}'$, are transmitted over an AWGN channel using BPSK.

At the receiver, in TLC approach, the LDPC code is decoded using BP to yield an estimate $\hat{\mathbf{b}}$ for the transmitted bitstream $\mathbf{b}$. In our proposed MSL approach, each short BCH codeword $\mathbf{c}_i'$ for $i=1,\ldots,q$ is independently decoded using OSD to obtain the estimate $\hat{\mathbf{b}}'_i$ for the segment ${\mathbf{b}}'_i$. Then, segment estimates form $\hat{\mathbf{b}}' = [\hat{\mathbf{b}}'_1,\ldots,\hat{\mathbf{b}}'_q]$ as the overall estimate of $\mathbf{b}$. These decoders are selected for optimal BLER performance with respect to coding schemes: LDPC with BP approaches capacity at moderate blocklengths \cite{Mahyar2019ShortCode}, while BCH with OSD approaches the NA bound at short blocklength \cite{yue2023efficient}.

We note that TLC and the proposed MSC approach exhibit distinct error behaviors. Specifically, TLC distributes redundancy globally, improving average performance but risking corruption of the entire message upon decoding failure.  In contrast, our proposed approach allows some blocks to be decoded successfully while others may fail.

After channel decoding, we convert $\hat{\mathbf{b}}'$ back to sentence estimate $\hat{\mathbf{s}}'$ using ASCII representation. Then $\hat{\mathbf{s}}'$ is fed into the proposed SEC module to fully recover the original sentence $\mathbf{s}$. The SEC module consists of two components: 

\begin{itemize}
    \item Tokenization: $\hat{\mathbf{s}}'$ is segmented into subword units transformed to tokens using byte-pair encoding (BPE) \cite{BPE}, which is a subword-based compression algorithm that merges frequent character pairs into tokens. BPE generates a token sequence $\hat{\mathbf{t}}' = [\hat{\mathbf{t}}'_1, \hat{\mathbf{t}}'_2, ..., \hat{\mathbf{t}}'_{z}]$ of length $z$. 
    \item BART Language Model: The token sequence $\hat{\mathbf{t}}'$ is then processed by a fine-tuned BART model that predicts the original sentence by leveraging contextual patterns. 
\end{itemize}

An example of the estimated original sentence $\mathbf{s}$ before and after SEC is shown in Table \ref{tab1}, using $q=6$ codewords of $(128,64)$ extended BCH codes, where each codeword carries $\frac{64}{8} = 8$ characters. The second codeword segment is initially decoded incorrectly but subsequently corrected by SEC.

We note that SEC can also be applied to the TLC approach, but its effectiveness is severely limited due to its decoding error behavior. This will be demonstrated in Section \ref{section5}.
\begin{table}[t]
\centering
\caption{Example of proposed MSC with SEC for $q=6$ with $(128,64)$ extended BCH codes, obtained at SNR = 2 dB}
\renewcommand{\arraystretch}{1}
\begin{tabular}{@{} p{2.8cm} p{5.5cm} @{}}
\toprule
\textbf{Original Sentence $\mathbf{s}$}  & Everything went back to normal in the town. \\
\midrule
\textbf{\shortstack[l]{Estimate $\mathbf{s}'$ before SEC}} & Everythi$\!\!\!\!\!\!\!\!\underbrace{\texttt{å↓Ò¼$◄$£ê}}_{\text{Decoding failure of $\hat{\mathbf{b}}_2'$}} \!\!\!\!\!\!\!\!$back to normal in the town. \\\midrule
\textbf{Estimate $\mathbf{s}''$ after SEC} & Everything went back to normal in the town. \\
\bottomrule
\end{tabular}
\label{tab1}
\vspace{-1.2em}
\end{table}

\vspace{-0.9em}
\subsection{Semantic Error Correction (SEC)}
\label{section3b}
\vspace{-0.3em}


In SEC, the tokenization process first transforms $\mathbf{s}'$ into a sequence of tokens via:
\begin{equation}
    \hat{\mathbf{t}}' = f_{\mathrm{BPE}}(\hat{\mathbf{s}}') = [\hat{\mathbf{t}}'_1, \hat{\mathbf{t}}'_2, ..., \hat{\mathbf{t}}'_z],
\end{equation}
where \(f_{\mathrm{BPE}}(\cdot)\) denotes the pretrained BPE tokenization function from \cite{BPE}. Each token $\hat{\mathbf{t}}'_i$ is a high-dimensional vector representing a subword unit from the sentence, which may be a complete word, a word fragment, or even a single character, depending on the tokenization process. BPE find subword units by iteratively merging adjacent characters according to pre-learned text statistics. The length $z$ varies across sentences depending on their semantic content. The initial estimate $\mathbf{s}'$ after channel decoding often contains corrupted text with garbled symbols or erroneous words. In such cases, BPE will recursively decompose the input into smaller subword units, eventually falling back to individual characters when no valid merges are applicable.

With tokens $\hat{\mathbf{t}}'$, SEC operates at the sentence semantic level, restoring linguistic coherence by leveraging the BART model
\begin{equation}
    \hat{\mathbf{s}}'' = f_{\mathrm{BART}}(\hat{\mathbf{t}}';\bm{\theta}),
\end{equation}
where $\hat{\mathbf{s}}''$ is the final estiamte of $\mathbf{s}$ and $f_{\mathrm{BART}}$ is the BART model with parameters $\bm{\theta}$. BART consists of an encoder and an autoregressive decoder. The BART encoder transforms  $\hat{\mathbf{t}}'$ into hidden representations $\mathbf{H}$, where each vector in $\mathbf{H}$ captures information from the entire sequence. The decoder then generates the output sequence token-by-token from left to right. At each step $i$ ($1\leq i \leq z$), it computes a conditional probability of each target token given by
\begin{equation} \label{equ::token:prob}
     P(\hat{t}''_i \mid \hat{t}''_{<i}, \mathbf{H})
\end{equation}
where $\hat{t}''_{<i} = (\hat{t}''_1, \ldots, \hat{t}''_{i-1})$ denotes the partial output token generated up to step \(i-1\). Then, it selects the token from vocabulary $\mathcal{V}$ with the highest probability as the next output token in the inference step, i.e., $$\hat{t}''_i = \arg\max\limits_{t \in \mathcal{V}}\ \ P(t \mid \hat{t}''_{<i}, \mathbf{H}).$$

Next, output tokens $\{\hat{t}''_1,\ldots,\hat{t}''_z\}$ are transform into the complete output sequence $\hat{\mathbf{s}}''$. Thus, The overall probability of generating $\hat{\mathbf{s}}''$ given the input $\hat{\mathbf{s}}'$ is the product of these conditional probabilities of tokens; that is
\begin{equation}
    P(\hat{\mathbf{s}}'' \mid \hat{\mathbf{s}}') = \prod_{i=1}^{z'} P(\hat{t}''_i \mid \hat{t}''_{<i}, \mathbf{H}),
\end{equation}
which will be used in the loss function for training.

Although BART is pretrained on language denoising tasks, it requires specific fine-tuning for channel decoding errors in our scheme. To fine-tune the model parameters $\bm{\theta}$, we construct a synthetic dataset 
\(\mathcal{D} = \{(s^{(j)}, \hat{s}'^{(j)})\}_{j=1}^{N}\) where $N$ represents the total number of sentence pairs. Each pair consists of an original sentence \(s^{(k)}\) and its corresponding estimate $\hat{s}'^{(k)}$ after channel decoding but before SEC, obtained through simulations over an AWGN channel at specific SNRs.

We fine-tune BART using two complementary loss functions. The first is the cross-entropy loss given by
\begin{equation}
    \mathcal{L}_{\text{seq}} = - \frac{1}{N} \sum_{j=1}^{N} \log P\left(\mathbf{s}^{(j)} \mid \hat{\mathbf{s}}'^{(j)}\right).
\end{equation}
This loss can promote semantic accuracy but may favor meaning preservation over exact sequence matching. The model may substitute tokens to maintain semantic integrity, resulting in outputs that are semantically correct but fail to match the original sequence exactly. To address this, we add the Levenshtein edit distance loss:
\begin{equation} \label{equ::Lloss}
    \mathcal{L}_{\text{edit}} = \frac{1}{N} \sum_{j=1}^{N} \frac{D\left( \hat{\mathbf{s}}'^{(j)}, \mathbf{s}^{(j)}\right)}{\ell_j + \delta},
\end{equation}
where $D(\cdot, \cdot)$ denotes the Levenshtein edit distance. $\ell_j$ is the character length of $\mathbf{s}^{(j)}$, and $\delta>0$ is a chosen hyperparameter. The Levenshtein distance measures the minimum number of single-character operations (insertions, deletions, or substitutions) required to transform $\hat{\mathbf{s}}'^{(j)}$ into $\mathbf{s}^{(j)}$.

Finally, our total loss function is given by
\begin{equation} \label{equ:::overallLoss}
    \mathcal{L}_{\text{total}} = \mathcal{L}_{\mathrm{seq}} + \alpha \cdot \mathcal{L}_{\mathrm{edit}},
\end{equation}
where \(\alpha\) is a tunable hyperparameter. The total loss function combines semantic understanding with sentence structural fidelity. We use the standard gradient-based training process with the Adam optimizer to update the parameters $\bm{\theta}$ for BART; the details are omitted for brevity.


\vspace{-0.8em}
\subsection{Semantic Confidence HARQ}
\label{section3c}
\vspace{-0.3em}
SEC can correct decoding errors of partial segments in most cases, but it may still fall short when confronted with several corrupted blocks that contain critical information. For example, if the noun ``antibiotics'' is lost due to decoding errors in ``The patient requires antibiotics.'' The SEC module may fail to recover the correct medication base on the surrounding context. To address this challenge, we introduce a semantic confidence-guided HARQ (SemConf HARQ) mechanism that identifies uncertain tokens, and selectively requests retransmission of codeword segments that cover critical uncertain tokens. 

Conventional HARQ typically uses a cyclic redundancy check (CRC) to detect decoding errors and request retransmission of entire messages, which can introduce error detection loss and the code rate loss with extra CRC bits. In contrast, our approach leverages the semantic confidence measures from the SEC to detect decoding errors and identify exactly which segments require retransmission. 
The key insight is that BART decoder generates each output token $\hat{t}''_i$ with an associated probability, i.e., $P(\hat{t}''_i \mid \hat{t}''_{<i}, \mathbf{H})$ given in \eqref{equ::token:prob}.
This probability reflects the certainty in generating \(\hat{t}_i''\). A higher probability indicates a higher confidence. 

There exists a mismatch between the token boundaries and codeword segment boundaries. A single token may span multiple codeword segments, and conversely, a single codeword segment may contain parts of multiple tokens. Thus, we define the confidence of the decoding result $\hat{\mathbf{b}}_j'$ from the $j$-th codeword segment as:
\begin{equation} \label{equ::conf:score}
  \gamma_j = \frac{1}{|\mathcal{T}_j|}\sum_{i \in \mathcal{T}_j} \omega_{i,j} \cdot  P(\hat{t}''_i \mid \hat{t}''_{<i}, \mathbf{H}),
\end{equation}
where $\mathcal{T}_j$ is the set of tokens whose corresponding subword units have any overlap with segment $\hat{\mathbf{b}}_j'$, and $\omega_{i,j}$ is the proportion of token $\hat{t}''_i$ that falls within $\hat{\mathbf{b}}_j'$, based on character count of the subword unit.

After computing confidences for all segments, i.e., $(\gamma_1,\gamma_2,\ldots,\gamma_q)$. We use a threshold $T \in (0,1)$ to determine decoding error of segments. If $\gamma_i < T$, the segment is considered potentially incorrect. Then, the $u$ codeword segments with the lowest confidence less than $T$ will be requested for retransmission. The values of $u$ and $T$ should be selected according to the system requirements to balance performance and retransmission overhead. Upon receiving the retransmitted segments, the receiver first performs channel decoding for these segments. Then, it replaces the corresponding sections in the overall sentence estimate $\mathbf{s}'$ and performs SEC again.

\vspace{-1em}
\section{Experimental Results and Discussion}
\vspace{-0.3em}
\label{section4}
\subsection{Experimental Setup} 
\vspace{-0.2em}
\subsubsection{Dataset}
We fine-tune the SEC and evaluate our proposed model using the Stanford Natural Language Inference (SNLI) corpus \cite{snli}, a widely adopted benchmark for sentence-level semantic understanding. From the corpus, we randomly selected 20,000 sentences for training and 500 for testing.

For training data, each of the 20,000 sentences is converted to its ASCII representation and encoded by a channel encoder. It is then transmitted through a simulated AWGN channel under a range of various SNRs, and decoded by a channel decoder to obtain its estimated sentence. This AWGN procedure is repeated 10 times for each sentence, resulting in a total of 200,000 original-decoded sentence pairs for model training. 

\subsubsection{Implementation Configuration}
Our experiments use sentences from the SNLI corpus with character lengths $\ell$ ranging from 57 to 64. To ensure consistent processing, we apply zero-padding during ASCII encoding when $\ell < 64$, creating fixed-length inputs of 512 bits (64 bytes) before channel coding.
We compare two channel coding schemes:

\begin{itemize}
    \item TLC Approach: We implement an LDPC code with $n=1048$ and $k=512$, decoded using BP with 18 iterations.
    \item Proposed MSC Approach:  We use the extended BCH code $\mathcal{C}(n',k')$ with parameters $n' = 128$ and $k' = 64$. Thus, each sentence is divided into $q=8$ equal segments for independent encoding and transmission. At the receiver, each BCH codeword is decoded by an order-4 OSD decoder to acheive near-MLD performance.
\end{itemize}

All transmissions, modulation, AWGN channel, channel encoders, and decoders are simulated using the Sionna library~\cite{sionna}.  For the SEC module, we fine-tune the pre-trained BART-base model from Huggingface Transformers \cite{HF}. All experiments are conducted on a machine equipped with an NVIDIA A10G GPU and an AMD EPYC 7R32 CPU. The detailed training parameters are summarized in Table~\ref{table:sim_params}.

\begin{table}[t]
\centering
\footnotesize
\caption{Training parameters}
\renewcommand{\arraystretch}{1}
\begin{tabular}{ll|ll}
\toprule
\textbf{Parameter} & \textbf{Value} & \textbf{Parameter} & \textbf{Value} \\
\midrule
Training SNRs & $-2$ to $2$ dB & Learning rate & $3 \times 10^{-5}$ \\
Optimizer & Adam & Batch size & 128 \\
$\alpha$ in \eqref{equ:::overallLoss} & 0.6 & $\delta$ in \eqref{equ::Lloss}& 0.0001\\
\bottomrule
\end{tabular}
\label{table:sim_params}
\vspace{-1.2em}
\end{table}

\begin{figure*}[htbp]
    \centering
    \includegraphics[width=0.9\textwidth]{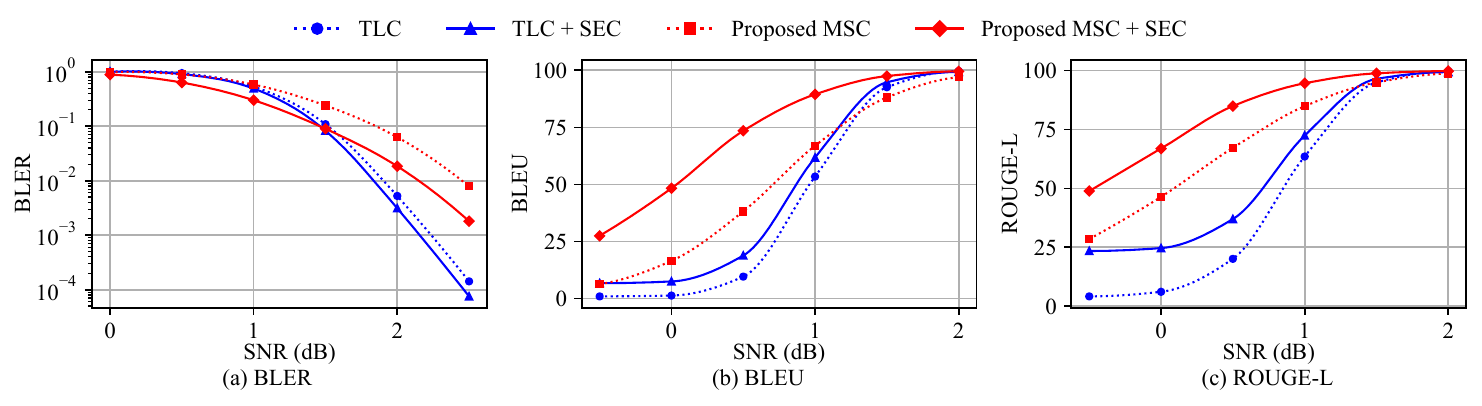}
    \vspace{-0.5em}
    \caption{Performance of proposed MSC scheme comapred to TLC scheme in BLER, BLEU, and ROUGE-L.}
    \label{fig3}
    \vspace{-1em}
\end{figure*}

\begin{figure*}[htbp]
    \centering
    \includegraphics[width=0.9\textwidth]{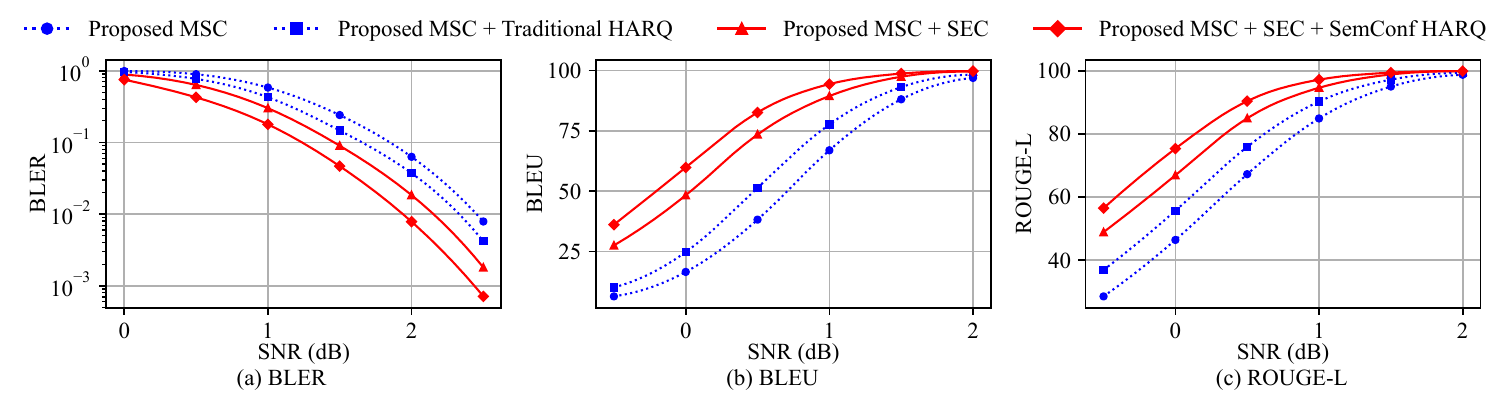}
    \vspace{-0.5em}
    \caption{Performance of proposed proposed MSC scheme with SemConf HARQ in BLER, BLEU, and ROUGE-L.}
    \label{fig4}
    \vspace{-1em}
\end{figure*}

\subsubsection{Evaluation metrics}
Our evaluation framework captures both traditional transmission reliability and semantic fidelity. 

For traditional transmission reliability, we evaluate BLER, which is defined as the ratio between the number of incorrectly received blocks and to total transmitted blocks. In this paper, each sentence $\mathbf{s}$  is considered as a single "block" when calculating BLER,  regardless of its encoding method. A block error occurs when the recovered sentence differs from the original one (by even a single character),  either directly after channel decoding ($\mathbf{s}\neq \mathbf{s}'$) or after additional SEC ($\mathbf{s}\neq \mathbf{s}''$). 

For semantic evaluation, we use two standard metrics from natural language processing: BLEU and ROUGE-L. In our work, BLEU quantifies text similarity by calculating the precision of matching word sequences (1-grams to 4-grams) between generated and reference texts \cite{bleu}. Higher BLEU scores indicate better preservation of the exact wording from the original message. ROUGE-L is based on the longest common subsequence (LCS) \cite{rouge}, which assesses global structural similarity and semantic content preservation.

\vspace{-1em}
\subsection{Performance of the proposed approach with SEC}
\label{section4c}
    \vspace{-0.4em}

In this section, we investigate the performance of proposed MSC with and without SEC. The TLC approach is selected as the benchmark. We also apply our fine-tuned SEC module to the TLC for a comprehensive comparison.

\subsubsection{BLER Performance}
The BLER performance is illustrated in Fig.~\ref{fig3}(a). At the low SNR range, both TLC and MSC schemes suffer from poor BLER performance. As the $\text{SNR}$ increases, the long code exhibits a pronounced waterfall region starting around $1~\text{dB}$, reaching a BLER of $10^{-4}$ at $2.5~\text{dB}$, demonstrating its strong error-correction capability. In contrast, the MCS scheme without SEC only reaches a BLER of $10^{-2}$ at $2.5~\text{dB}$. This is because of the channel distortion induced by the finite blocklength bound.

The introduction of the SEC module reveals different effects for the two coding strategies. For TLC, SEC yields only a modest BLER improvement at high SNRs. In contrast, the proposed MSC scheme benefits significantly from SEC, which provides a performance gain of approximately $0.5~\text{dB}$ in BLER the case without SEC at all SNRs. With the SEC, the short block code approach even outperforms the long code scheme in BLER at low SNRs. This demonstrates the effectiveness of combining MSC with SEC.

\subsubsection{BLEU and ROUGE-L Score}
Figures~\ref{fig3} (b) and (c) shows that the proposed MSC scheme with SEC consistently achieves the highest BLEU and ROUGE-L scores across the full SNR range. Notably, at $\text{SNR} = -0.5~\text{dB}$, SEC improves BLEU from $6$ to $28$, and ROUGE-L from $5$ to $48$ compared to the MSC without SEC. However, SEC only improves the semantic scores of TLC slightly. The effectiveness of SEC for MSC is largely due to the localized nature of errors in multiple short code blocks, which enables it to utilize contextual information more efficiently. This also explains why the MSC without SEC also outperforms the TLC at low SNRs in terms of BLEU and ROUGE-L. We note that even at high SNRs where the long code begins to have ``waterfall'' BLER performance, our proposed MSC with SEC delivers the best semantic fidelity. 

\begin{table}[t]
\centering
\caption{Performance comparison of various schemes at SNR = 1 dB.}
\renewcommand{\arraystretch}{1}
\begin{tabular}{l|c|c|c}
\toprule
\textbf{Scheme} & \textbf{BLER} & \textbf{BLEU} & \textbf{ROUGE-L} \\
\midrule
TLC                    & 0.5333 & 53.45 & 63.52 \\
TLC + SEC              & 0.4913 & 61.73 & 72.41 \\
MSC ($q=16$)           & 0.9337 & 43.28 & 74.24 \\
MSC ($q=16$) + SEC     & 0.5490 & 82.58 & 91.72 \\
MSC ($q=8$)            & 0.5828 & 66.96 & 84.90 \\
MSC ($q=8$) + SEC      & 0.3002 & 89.46 & 94.63 \\
\bottomrule
\end{tabular}
\label{tab:shortcode}
\vspace{-2em}
\end{table}

\subsubsection{The impact of Segment Number $q$}
In the proposed MSC, when fixing the overall encoded sequence length $n$, the selection the number of segments $q$ poses an important tradeoff. A larger $q$ provides more opportunities for semantic context clues through additional short codeword segments. However, each segment will have a smaller blocklength, resulting in lower decoding latency but degraded individual error performance. 

To evaluate the performance of MSC approach with different $q$, we keep the same setting of TLC, but further adopt a shorter code length with $\mathcal{C}(n'=64,k'=32)$ polar code in MSC. Accordingly, each sentence is divided into $q=16$ segments. Note that there is no extended BCH available at this $k'$ and $n'$. Thus, we selected polar codes that also approach the NA bound. Each polar codeword is decoded by order-3 OSD to achieve near-MLD performance. One can use sequential decoding to achieve a lower decoding latency \cite{yue2023efficient}.

Table~\ref{tab:shortcode} compares the performance of TLC, MSC ($q=8$), and MSC ($q=16$) at SNR of 1~dB. As shown, MSC ($q=16$) without SEC exhibits worst performance across all three metrics. After applying SEC, its performance is significantly improved, achieving a similar BLER performance to the MSC ($q=8$) without SEC, while showing better BLUE and ROUGE-L scores. However, MSC ($q=16$) with SEC is still worse than MSC ($q=8$) with SEC across all metrics.

\subsubsection{Latency Analysis}
We analyze the receiving latency performance by measuring both the time consumption of channel decoding and the SEC module.  For our proposed MSC, we assume that the channel decoding of multiple short block code segments is performed in parallel.

The latency breakdown for each scheme is summarized in Table~\ref{tab:latency}. Specifically, TLC employing LDPC with BP decoding requires approximately 400\, ms per sentence. In contrast, MSC ($q=8$) using BCH code with OSD decoding reduces decoding latency significantly to around 160\, ms per sentence. Furthermore, MSC ($q=16$) with polar codes further reduces the decoding latency to around 87 ms.

In our implementation, the delay introduced by the SEC module is 80\, ms. Therefore, the total latency for TLC with SEC reaches approximately 480\, ms, whereas MSC ($q=8$) with SEC achieves a notably lower total latency of approximately 240\, ms, highlighting its advantage in receiving latency. MSC ($q=16$) achieves the lowest latency of approximately 167\, ms, but at the cost of degraded recovery performance.

\vspace{-1.1em}
\subsection{Performance of the proposed SemConf HARQ}
\vspace{-0.2em}
We further evaluate the performance of the proposed SemConf HARQ mechanism, as shown in Fig.~\ref{fig4}. For fair comparison, we consider a baseline scheme legend as ``MSC + traditional HARQ,'' in Fig.~\ref{fig4}. This scheme is not aided by SEC and uses the traditional CRC method to determine unsuccessfully decoded segments for retransmission. For simplicity, we omit the additional CRC bits and assume perfect CRC accuracy. Note that practical CRC will introduce error-detection loss and code rate loss, resulting in even worse throughput and reliability performance. 

For all comparisons in this subsection, we use $q=8$ segments in the proposed MSC. We assume each scheme is limited to a single retransmission round of one of the detected erroneous segments. For SemConf HARQ, this retransmitted segment is selected according to the confidence score given by \eqref{equ::conf:score}, while the traditional HARQ baseline randomly requests one erroneous segment for retransmission. We also include the proposed MSC without HARQ for comprehensive comparison.

The BLER performance is compared in Fig.~\ref{fig4}(a). As can be seen, the traditional HARQ approach yields a slight gain over the MSC without HARQ and SEC. SemConf HARQ achieves the best BLER performance, offering approximately a $0.7~\text{dB}$ gain over the MSC without SEC, a $0.5~\text{dB}$ improvement over the traditional HARQ, and a $0.25~\text{dB}$ advantage over the MSC with SEC configuration. These results highlight the efficacy of integrating the semantic confidence level into HARQ.

Fig.~\ref{fig4}(b) and \ref{fig4}(c) compare the semantic metrics, further underscoring the benefit of SemConf HARQ. Traditional HARQ, which lacks semantic processing, provides only marginal gains in BLEU and ROUGE-L metrics. In contrast, the proposed SemConf HARQ framework significantly improves semantic fidelity, achieving nearly a threefold increase in BLEU scores and a $60\%$ gain in ROUGE-L at $\text{SNR} = 0~\text{dB}$, compared to the baseline MSC without SEC. Furthermore, SemConf HARQ achieves more than $0.2~\text{dB}$ improvement in semantic-level performance compared to the MSC with SEC. We highlight that SemConf HARQ with additional transmission of only one $(64,128)$ codeword block, i.e, $\frac{1}{8}$ of the overall stream of $1024$ bits, can significantly increase the BLER and semantic performance. Thus, it can substantially reduce retransmission overhead and delay compared to traditional HARQ methods.

\begin{table}[t]
\centering
\caption{Latency breakdown for each decoding scheme (per sentence).}
\renewcommand{\arraystretch}{1}
\begin{tabular}{ll|ll}
\toprule
\textbf{Scheme} & \textbf{Time(ms)} & \textbf{Scheme} & \textbf{Time(ms)} \\
\midrule
TLC & 400 & TLC + SEC & 480 \\
MSC ($q=8$) & 160 & MSC ($q=8$) + SEC & 240 \\
MSC ($q=16$) & 87 & MSC ($q=16$) + SEC & 167 \\
\bottomrule
\end{tabular}
\label{tab:latency}
\vspace{-1.5em}
\end{table}

\vspace{-1em}
\section{CONCLUSION}
\vspace{-0.4em}
\label{section5}
This paper presented a novel semantic-enhanced decoding scheme for transmitting natural language sentences with multiple short block codes. At the receiver, we proposed a semantic error correction (SEC) module based on BART to recover the transmitted sentence through context reasoning. Simulation results show that this proposed scheme outperforms transmission with a single long code in terms of block error rate and semantic similarity metrics. Furthermore, we proposed a semantic HARQ mechanism, which requests re-transmission of selective codeword segments based on their semantic confidence. This method can significantly enhance the BLER and semantic performance with only one additional retransmission of a single short block.


\ifCLASSOPTIONcaptionsoff
\newpage
\fi

\vspace{-1em}
\bibliographystyle{IEEEtran}
\bibliography{IEEEabrv, refs}
\vspace{-1em}

\end{document}